\documentclass[aps,prl,reprint,superscriptaddress,tightenlines,amsmath,amssymb]{revtex4-1}

\usepackage{siunitx}
\usepackage{graphicx}
\usepackage{mathptmx}
\usepackage[colorlinks=true,urlcolor=blue,linkcolor=blue,citecolor=blue]{hyperref}
\DeclareSIUnit\molar{\textsc{M}}
\DeclareSIUnit\pixel{px}
\DeclareSymbolFontAlphabet{\mathcal}{symbols}
\DeclareSymbolFont{symbols}{OMS}{cmsy}{m}{n}
\begin{document}

\title{Morphology and stability of droplets sliding on soft viscoelastic substrates.}

\author{Mathieu Ol\'{e}ron}
\author{Laurent Limat}
\author{Julien Dervaux}
\author{Matthieu Roch\'{e}}
\email[E-mail address: ]{matthieu.roche@u-paris.fr}

\affiliation{Universit\'{e} Paris Cité, Mati\`{e}re et Syst\`{e}mes Complexes, CNRS UMR 7057, F-75013 Paris, France, EU}

\begin{abstract}
  We show that energy dissipation partition between a liquid and a solid controls the shape and stability of droplets sliding on viscoelastic gels. When both phases dissipate energy equally, droplet dynamics is similar to that on rigid solids. When only the solid dissipate, we observe an apparent contact angle hysteresis, of viscoelastic origin. We find excellent agreement between our data and a non-linear model of the wetting of gels of our own that also indicates the presence of significant slip. Our work opens general questions on the dynamics of curved contact lines on compliant substrates.
\end{abstract}

\maketitle

An ever-increasing number of applications such as biofouling repellency \cite{maccallum2015,lavielle2021}, dew harvesting \cite{Sokuler2010,phadnis2017,sharma2022} and anti-icing \cite{mittal2020} relies on the use of viscoelastic coatings. Not only do these materials confer controlled interfacial properties to their substrate, they may be compliant enough to deform and dissipate energy after the deposition of a droplet \cite{carre1996,long1996a,style2013c,Park14}, affecting both the equilibrium shape of the latter and the dynamics of the contact line between the solid, the liquid and the ambient fluid \cite{carre2001,karpitschka2015,dervaux2015}. This coupling leads to unique substrate-droplet and droplet-droplet interactions \cite{Style13b,karpitschka2016,zhao2018,smith-mannschott2021,khattak2022}.

\begin{figure}[!ht]
  \centering
  \includegraphics[scale=1]{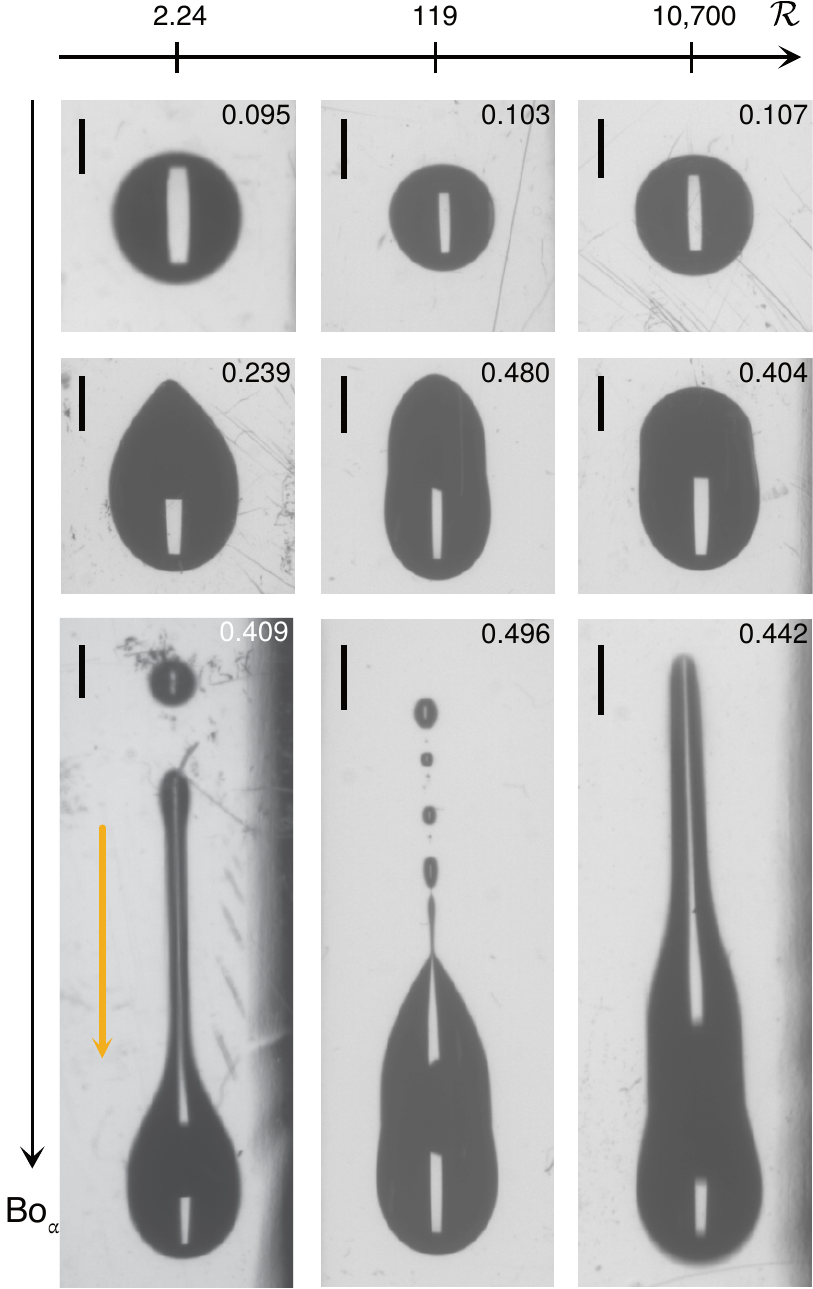}
     \caption{Top view of droplets sliding on a silicone gel as a function of the Bond number $Bo_{\alpha}$ and the relaxation ratio $\mathcal{R}$. Each image displays the value of $Bo_{\alpha}$. The pinch behind the front of the droplet at large $\mathcal{R}$ is an artefact: the equilibrium contact angle is greater than $\pi/2$ for these systems and the liquid/air interface hangs over the moving contact line. Orange arrow: direction of motion. Scale bar: $2$ \si{\milli\meter}.}
     \label{fig:shapes}
\end{figure}
The influence of substrate compliance on the shape and stability of moving droplets is hardly known. The morphology of droplets sliding down a rigid plate inclined at an angle $\alpha$ with the horizontal is dictated by the dependence of the dynamic contact angle $\theta_{d}$ between the liquid-gas and solid-liquid interfaces on droplet velocity $U$ and the ability of the trailing edge contact line to form corners \cite{podgorski2001,rio2005,snoeijer2005,legrand2005,peters2009,winkels2011}. Beyond a threshold velocity, corners destabilize into rivulets that fragment into tinier droplets known as pearls \cite{podgorski2001,rio2005,legrand2005,puthenveettil2013,kim2015,liang2017}. These results hold for systems where energy dissipation occurs entirely in the liquid. However, on a gel with surface energy $\gamma_s$ and shear modulus $\mu_0$, deformations having a magnitude comparable to the elastocapillary length $\ell_s = \gamma_s/(2\mu_0)$ propagate with the contact line, leading to dissipation in the droplet and the substrate and a different $\theta_d(U)$ relationship. Thus, we expect a more complex shape selection process.

\begin{figure*}[!t]
  \centering
  \includegraphics[scale=1]{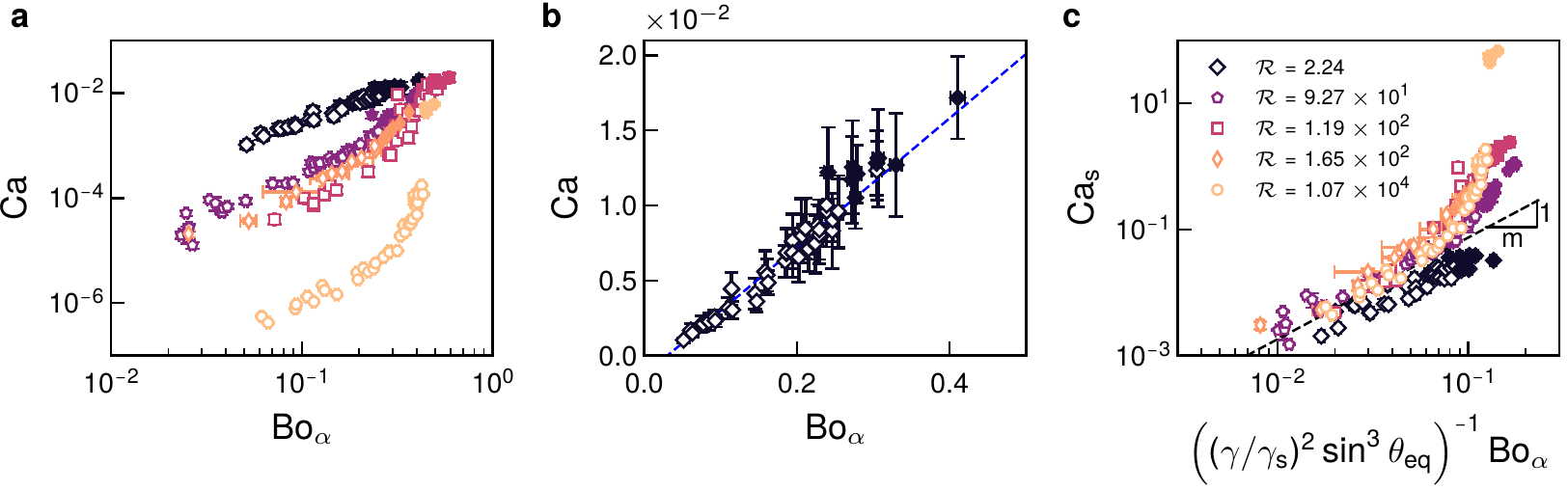}
     \caption{(a) Dependence of the liquid capillary number $Ca$ on the Bond number $Bo_{\alpha}$. (b) Focus on the case $\mathcal{R} \simeq 2$. Blue dashed line: Eq.\,\ref{eq:liquid_scaling}. (c) Dependence of the solid capillary number $Ca_s$ on the Bond number $Bo_{\alpha}$. Dashed line: Eq.\,\ref{eq:solid_scaling}.  In all panels, filled symbols correspond to points measured above the pearling threshold. (a) and (c) share the same legend.}
     \label{fig:Bo}
\end{figure*}
Here we tune energy dissipation partition between the liquid and the solid and show that this balance controls the morphology and stability of droplets sliding on soft solids. When both materials dissipate energy equally, droplet dynamics resemble those reported on a rigid substrate, despite gel compliance. When dissipation occurs only in the substrate, we observe an apparent contact angle hysteresis and the appearance of corners only beyond the pearling instability threshold. We rationalize our results with a non-linear model of the wetting of soft solids of our own. The very good agreement we obtain brings strong evidence that the apparent hysteresis originates from the viscoelastic force exerted by the solid on the moving contact line. Our analysis also highlights the existence of significant slip due to the presence of free polymer chains in the gel. Finally we discuss the perspectives of our work regarding the role of deformable substrates in setting the properties of curved contact lines.


 We perform experiments on a silicone gel with shear modulus $\mu_0\simeq 1$ kPa, relaxation time $\tau=18.2$ ms and a 62-wt\% free chain content. We use Newtonian liquids with surface tensions $\gamma$, densities $\rho$ and viscosities $\eta$. Details about the materials and protocols can be found in the Supplemental Materials \cite{suppMat}. Sliding experiments are characterized by two dimensionless numbers, the Bond number $Bo_{\alpha}=\rho g R_0^3\sin{\alpha}/(\gamma R_c)$ and the liquid capillary number $Ca=\eta U/\gamma$, that compare capillary stresses to  gravitational and viscous ones respectively. Here, $g$ is the acceleration of gravity, $R_0$ is the radius of the spherical droplet before deposition, and $R_c$ is the contact radius between the droplet and the substrate. Energy dissipation partition between the two media is characterized by the relaxation ratio $\mathcal{R}$ \cite{karpitschka2015,dervaux2020} that compares the viscocapillary relaxation velocity in the liquid, $U_{ l} = \gamma/\eta$, to that in the solid, $U_{s} = \ell_s/\tau$, giving $\mathcal{R}=\gamma\tau/(\eta\ell_s)$. Contact line motion has been studied in the limit $\mathcal{R}\rightarrow\infty$ in the literature \cite{Style2017,andreotti2020}, \textit{i.e.}\,energy dissipation in the liquid is neglected. The liquids we use allow us to vary $\mathcal{R}$ over four orders of magnitude.


Droplet shapes carry obvious signatures of changes in the magnitude of $\mathcal{R}$ (Fig.\,\ref{fig:shapes}). While droplets remain nearly axisymmetric at small Bond numbers, symmetry is lost as $Bo_{\alpha}$ increases. When $\mathcal{R} \simeq 2$, a corner appears at the trailing edge. In contrast, the aft and fore radii of curvature of the droplets are comparable when $10^2\leq\mathcal{R}\leq10^4$. Besides, the droplet contour contains portions parallel to the direction of motion, leading to shapes similar to those observed in the case of droplets sliding on hysteretic surfaces \cite{dussanv.1983,podgorski2000,puthenveettil2013,ahmed2014a}. A further increase of $Bo_{\alpha}$ at all values of $\mathcal{R}$ leads to the observation of the pearling instability.

Figure \ref{fig:Bo}a shows that, for equivalent Bond numbers, liquid capillary numbers vary over four orders of magnitude as the relaxation ratio changes by the same amount. The data for $\mathcal{R}\simeq2$ suggest an affine relation between $Ca$ and $Bo_{\alpha}$ (Fig.\ref{fig:Bo}b), with a non-zero y-intercept, similar to the rigid case \cite{podgorski2001,legrand2005}. The functional form for the other datasets is more complex. We multiply $\mathcal{R}$ with $Ca$ to obtain a capillary number for the solid, $Ca_s=U\tau/\ell_s$ and plot the data in Figs.\,\ref{fig:Bo}a-b as a function of this quantity. Accounting for variations of the equilibrium contact angle $\theta_{eq}$ from one system to another \cite{suppMat}, we observe a collapse of the large-$\mathcal{R}$ data  on a single master curve (Fig.\,\ref{fig:Bo}c). We can discriminate the curve obtained for $\mathcal{R}\simeq 2$, in line with the assumption that the power balance relevant to these experiments differs from the one tested in this representation. 

\begin{figure*}[!ht]
  \centering
  \includegraphics[scale=1]{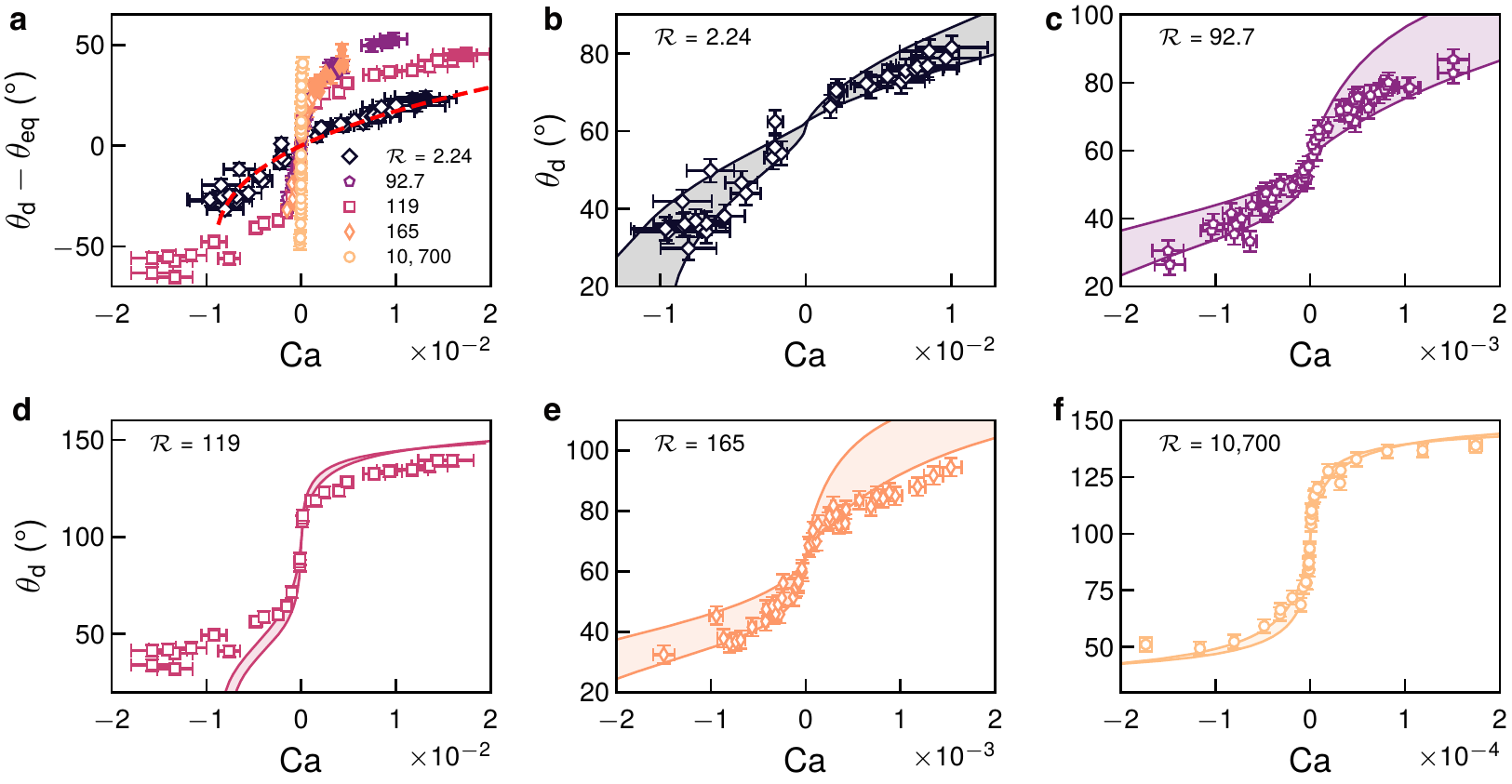}
  \caption{(a) Dependence of the deviation from the equilibrium contact angle $\theta_d-\theta_{eq}$ on the liquid capillary number. Positive (resp.\,negative) $Ca$ values correspond to leading-edge (resp.\,trailing-edge) angles, and advancing (resp.\,receding) contact lines. Filled symbols: points measured above the pearling threshold. Red dashed line : fit of the Cox-Voinov law (Eq.\,\ref{eq:Cox-Voinov}) to the $\mathcal{R} \simeq 2$ data. (b)-(f) Comparison between experimental dynamic contact angles $\theta_d$ and Eq.\,\ref{eq:fitDervaux}. The fitting parameter is the microscopic cutoff length scale, $\lambda=100$ nm for all systems. Colored areas: fit spread reflecting experimental uncertainties.}
  \label{fig:NLM}
\end{figure*}

Another way to characterize the dynamics of sliding droplets is to measure the dependence of the dynamic contact angle on droplet velocity. Figure \ref{fig:NLM} shows the deviation from the equilibrium contact angle $\theta_{eq}$ \cite{suppMat} as a function of the capillary number $Ca$, for each system. We observe that the apparent dynamic contact angle $\theta_{d}$ increases smoothly as the capillary number goes from negative to positive values when $\mathcal{R}\simeq 2$. Corners appear when $Ca\geq 7.5\times 10^{-3}$, a value of the same order of magnitude as those reported for fluoropolymer-coated silicon wafers \cite{podgorski2001,rio2005,legrand2005}. The other curves display a steep jump of several tens of degrees around $Ca=0$ that brings to mind results obtained in the case of significant wetting hysteresis \cite{eral2013} and when a contact line moves on low-modulus natural rubber and \textit{cis}-butadiene \cite{extrand1997}. In a vein similar to what we observed in figure \ref{fig:Bo}, the datasets collapse on a master curve when plotted against the solid capillary number $Ca_s$ \cite{suppMat}. The curve obtained at $\mathcal{R}\sim10^4$ displays plateaus in the advancing and receding branches, similar to those reported in earlier studies \cite{karpitschka2015}.

The data presented in figure \ref{fig:Bo} can be discussed in terms of scaling laws. We focus first on the case $\mathcal{R}\simeq 2$, and we follow a rationale proposed in studies of droplets sliding on rigid substrates \cite{podgorski2001,legrand2005}. We assume that dissipation occurs entirely in the liquid and balances the gravitational force experienced by the droplet. Contact-angle hysteresis may also be present. The resulting force balance leads to \cite{suppMat}:
\begin{equation}
    Ca \sim \frac{1-\cos \theta_{eq}}{\sin \theta_{ eq}} [Bo_{\alpha} - Bo_c].
    \label{eq:liquid_scaling}
\end{equation}
where $Bo_c$ is a threshold Bond number capturing hysteresis effects. Scaling \ref{eq:liquid_scaling} captures well the  trend of the data obtained for $\mathcal{R} \simeq 2$ (Fig.\,\ref{fig:Bo}c). We can estimate the magnitude of the contact angle hysteresis of silicone gels from $Bo_c$ \cite{legrand2005,dussanv.1985a}  and we find $\Delta\theta=\theta_a-\theta_r\simeq 3.5$ \si{\degree}, a value compatible with the data displayed in Fig.\,\ref{fig:NLM}, the rounded shapes of droplets at $\mathcal{R}\simeq 2$ (Fig.\,\ref{fig:shapes}) and other reports in the literature regarding silicone gels \cite{maccallum2015,zhao2018}.

In the limit $\mathcal{R}\rightarrow\infty$, energy dissipates in the substrate. A reasoning similar to the one used in the previous paragraph leads to a scaling that balances  viscous dissipation in the solid and droplet weight \cite{suppMat}: 
 \begin{equation}
{Bo_{\alpha}} \sim \left( \frac{\gamma}{\gamma_{ s}} \right)^2 (\sin{\theta_{eq}})^3 {Ca_s}^m,
\label{eq:solid_scaling}
\end{equation}
where $m$ is the exponent of the power law describing the loss modulus of the substrate as a function of strain frequency \cite{suppMat}. For systems where $\mathcal{R} \geq 10^2$, Eq.\,\ref{eq:solid_scaling} captures our data up to $Bo_{\alpha}\simeq 8\times 10^{-2}$ (Fig.\,\ref{fig:Bo}c).


Given the similarities between the $\mathcal{R}\simeq2$ data and the rigid case, we compare the contact angle dependence on the capillary number to the Cox-Voinov law \cite{voinov1976,cox1986,legrand2005}:
\begin{equation}
    \theta_{d}^3 - \theta_{eq}^3 = 9 Ca \ln \left( \frac{h}{\lambda} \right),
    \label{eq:Cox-Voinov}
\end{equation}
where $h$ is the height on the liquid/vapor interface at which the angle is measured and $\lambda$ is a nanoscopic length scale introduced to circumvent stress divergence at the contact line. The agreement is qualitatively excellent (dashed line in Fig.\,\ref{fig:NLM}), provided the logarithmic term has an amplitude around 15. As we measure the contact angle at $h\sim 100$ \si{\micro\meter}, we obtain an unreasonable cutoff length scale $\lambda \simeq 30$ \si{\pico\meter}, smaller than an interatomic bond. The large value of the logarithmic term likely results from the fact that $\mathcal{R}\sim 1$: dissipation in the solid is of the same order of magnitude as in the liquid. Failure of the Cox-Voinov law is then expected, as it does not account for all dissipation sources.

We compare the data in Fig.\,\ref{fig:NLM} to a model that we proposed recently \cite{dervaux2020} that describes the wetting of soft solids in a regime of finite strains under the assumption that $\gamma_s$ is independent of strain: there is no Shuttleworth effect \cite{shuttleworth1950}. The model provides a prediction for the dependence of $\theta_{ d}$ on $Ca$ and $\mathcal{R}$ :
\begin{multline}
    g(\theta_d) = g \left( \frac{\pi}{2} + \arctan {\sqrt{\frac{\sqrt{1+\mathcal{A}^2(\mathcal{R},{Ca},\Lambda)}-1}{2}}}\right)\\
    + {Ca} \ln \left( \frac{h}{\lambda} \right),
    \label{eq:fitDervaux}
\end{multline}
with $g(x)=\int_0^x\frac{z-\sin{(z)}\cos{(z)}}{2\cos{(z)}}dz$, $\mathcal{A}$ a function that accounts for the capillary and viscoelastic forces at the contact line, and $\Lambda$ the ratio between the thickness of the substrate and the elastocapillary length $\ell_s$; here $\Lambda\rightarrow\infty$. This equation is formally similar to the general form of the Cox-Voinov relation, where the first r.h.s.\,term is related to the microscopic contact angle and the second results from dissipation in the liquid \cite{bonn2009}. Here, the microscopic angle term is a dynamic quantity set by the response of the solid to the  propagation of the ridge. The correction to this term increases as $\mathcal{R}$ increases.

Figure \ref{fig:NLM}b-f show that the agreement between the experimental data of Fig.\,\ref{fig:NLM}a and Eq.\,\ref{eq:fitDervaux}, setting $\lambda=$ 100 nm, is good to excellent for all datasets. In the case $\mathcal{R}=119$, the model fails to capture the receding branch $Ca<0$. The trailing edge of these droplets oscillate, and our model does not predict this response. The steepness of the $\mathcal{R}>>1$ curves close to $Ca =0$ is characteristic of soft hysteresis \cite{dervaux2020}. This hysteresis is apparent as contact angles remain defined at all velocities. However, its signature on droplet shape is akin to that of real hysteresis (Fig.\,\ref{fig:shapes}). These results and their interpretation clarify the nature of the hysteresis reported for soft materials in the past literature \cite{extrand1997}. They highlight the necessity to reach the smallest possible velocities to characterize wetting dynamics on soft solids accurately. Sliding droplets are a good system in this respect, as the sign of velocity near the contact line switches from positive to negative continuously along their contour \cite{rio2005,snoeijer2005}.

The fitting procedure also suggests that all systems, even at small values of $\mathcal{R}$, exhibit a steep asymptote at ${Ca}=0$. This is because viscoelastic dissipation in the solid, $\propto U^m$ with $m<1$, always exceeds viscous dissipation in the liquid, $\propto U$, at vanishing $Ca$. We note that the value of $\Delta\theta$ obtained from Eq.\,\ref{eq:liquid_scaling} is compatible with the jump magnitude seen in the fit of the model to the data for $\mathcal{R}\simeq 2$, indicating again that signaturess of soft hysteresis are akin to those of real, defect-induced, hysteresis. The model can also predict the full range of the data displayed in Fig.\,\ref{fig:Bo} \cite{suppMat}. 

Finally, fits to the datasets with Eq.\,\ref{eq:fitDervaux} are obtained while keeping $\lambda$ constant. We can also set experimental parameters to their nominal value an leave $\lambda$ free to adjust. We find best-fit values  $50\leq \lambda\leq 1000$ nm, larger than the molecular sizes of the liquids we use, of the order of 10 nm at most, and smaller than the elastocapillary length $\ell_{s}\simeq 20$ \si{\micro\meter}. The large magnitude of $\lambda$ is likely the result of the presence of free chains that are known to lubricate the contact between the droplet and the gel and induce significant slip \cite{lavielle2021,cai2022a,hauer2023}.

\begin{figure}[!ht]
    \centering
    \includegraphics[scale=1]{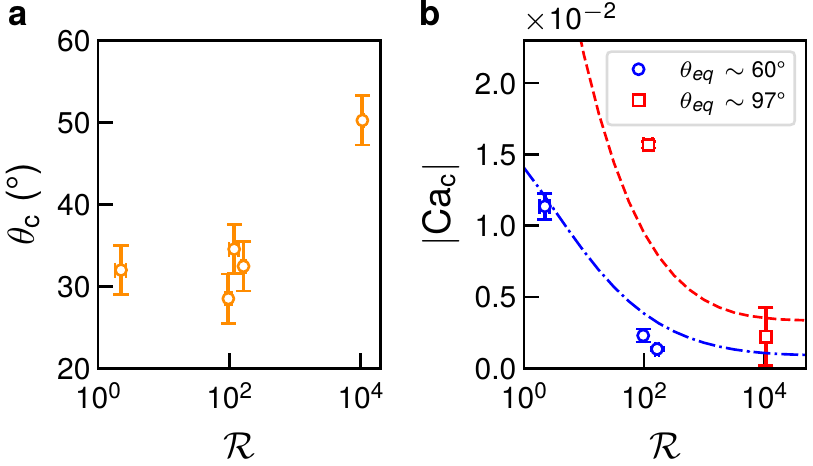}
       \caption{(a) Receding contact angle $\theta_{c}$ and (b) capillary number $Ca_{c}$ at the pearling threshold as a function of the relaxation ratio $\mathcal{R}$. $\theta_{c}$: receding contact angle of the last stable point; ${Ca_{c}}$: mean of the capillary numbers of the last stable and the first unstable points. Error on the latter is estimated with its standard deviation. Dashed lines: prediction of the model \cite{dervaux2020} for the vanishing-contact-angle capillary number.}
       \label{fig:instab}
\end{figure}

The characterization of the properties of the system at the threshold of the pearling instability shows how different the transition to liquid deposition on soft substrates is from that on rigid ones. First, the contact angle just before pearling, $\theta_{c}$, is around $32$\si{\degree} when the relaxation ratio $\mathcal{R} \lesssim 10^2$ and rises up to $50$\si{\degree} when $\mathcal{R} \simeq 10^4$ (Fig.\,\ref{fig:instab}a). These values are much larger than those, around \SI{10}{\degree}, reported for rigid substrates \cite{legrand2005}. Second, the threshold capillary number $\vert Ca_c \vert$ depends on both the equilibrium contact angle $\theta_{eq}$ and  $\mathcal{R}$ (Fig.\,\ref{fig:instab}b). While the former is expected \cite{snoeijer2007a} and may at least partly explain the jump of around an order of magnitude in $\vert Ca_c\vert$ that we see for $\mathcal{R}\sim 100$, the latter remains to be investigated. For $\mathcal{R}\simeq 2$,  $\vert Ca_c \vert \simeq 1.1\times 10^{-2}$ and it is around twice as large as in the rigid case for a comparable equilibrium contact angle $\theta_{eq}$ \cite{legrand2005}. Keeping the latter constant, a one-hundredfold increase of $\mathcal{R}$ decreases $\vert Ca_c\vert$ tenfold. This result stems from the fact that receding branches in the $\theta_d(Ca)$ curves tend to be steep, and their steepness increases with an increase of $\mathcal{R}$ \cite{dervaux2020}. As a consequence, for a constant equilibrium contact angle, the capillary number at which a zero receding contact angle is attained decreases as $\mathcal{R}$ increases. Thus, we expect the rivulet transition to occur at smaller values of $Ca$ for larger values of $\mathcal{R}$. This rationale explains why the prediction of our model for the zero-contact-angle receding capillary number captures quite well the data (Fig.\,\ref{fig:instab}b). It also suggests that dissipation in the substrate increases the sensitivity of droplets to the pearling instability.

In conclusion, we document how droplets slide on soft viscoelastic gels as a function of energy dissipation partition between the liquid and the substrate. While the substrate is always deformable, sliding droplets display shapes and dynamics akin to those observed on a rigid solid when dissipation occurs equally in both materials. When the substrate is the main dissipative element, straight lines parallel to the direction of motion appear in the droplet contour as the signature of an apparent hysteresis that appears when the dependence of dynamic contact angles on velocity is measured. A non-linear model that we proposed recently is able to describe the data very well. Thus, accounting for geometrical non-linearities is enough to describe droplet dynamics on soft substrates, without the need to assume a dependence of the surface energy of the solid on strain, in line with recent experimental results \cite{bain2021}. The model accounts at least qualitatively for the lubricating effect of free chains present in our system.

This study raises questions regarding the physics of curved contact lines. On rigid substrates, the trailing edge contact angle decreases almost to zero at the pearling transition threshold. The path to fragmentation is different on a soft gel. Our observations suggest that the curvature of the trailing edge is constrained by the substrate, an issue that calls for future work. In particular, contributions of the dry cross-linked network and the free chains should be controlled separately to investigate how elasticity \textit{via} the elastocapillary length and viscous dissipation \textit{via} the slip length set the shape of curved receding contact lines. The tenfold increase of the capillary number when passing the pearling threshold at $\mathcal{R}\simeq 10^4$ also deserves investigation, as we were not able to find a smooth transition. 

\begin{acknowledgments}
We thank L.\,Bergougnoux and E.\,Guazzelli for providing UCON U90. We are grateful to H.\,Mont\`{e}s for performing the time-temperature rheology of Sylgard 527. ANR (Agence Nationale de la Recherche) and CGI (Commissariat à l’Investissement d’Avenir) are gratefully acknowledged for their financial support through the GelWet grant (ANR-17CE30-0016).
\end{acknowledgments}

\section{Supplemental materials}
\subsection{Materials and Methods}
\subsubsection*{Liquids}

We use pure glycerol (G100, Sigma Aldrich, G5516), a 60-\%-glycerol-in-water mixture (G60), polyethylene glycol-ran-propylene glycol (P25, PEG-ran-PPG, average molecular weight $M_w \sim 2500$ \si{\gram\per\mole}, Sigma Aldrich), a 70\%-polyethylene glycol-ran-propylene glycol monobutyl ether-in-water mixture (P7, PEG-ran-PPG ME, Sigma Aldrich), and the UCON lubricant 75-H-90,000 (U90, Dow corning). Table \ref{tab:liquid_properties} summarizes their properties. All the liquids are insoluble in silicone gels. We measure the liquid-vapor surface tension $\gamma$ with the pendant drop technique. We estimate density $\rho$ by weighing a volume $V = 10\pm0.5$ mL of liquid with a 0.01-\si{\gram}-accurate scale. We measure the dynamic viscosity $\eta$ with a capillary viscosimeter sitting next to the set-up twice a day to account for hygroscopic and thermal effects. 
\begin{table}[!htb]
    \small
    \centering
    \caption{\small Properties of liquids used in our experiments.}
    \begin{tabular}{|l | c |c | c | c | c |c|}
    \hline
    \centering
           & Surface        &   Viscosity     &    Density      & Equilibrium    & Relaxation    \\
           & tension        &                 &                 & contact angle  & ratio         \\
           & $\gamma$       &    $\eta$       &       $\rho$    & $\theta_{ eq}$ & $\mathcal{R}$ \\
           & mN m$^{-1}$    &  mPa s          & $10^3$ kg m$^{-3}$ & \si{\degree}   & \\
    \hline 
    U90    &    40.7        &     36900       &       1.08      &    61.6        &    2.24        \\
           &   $\pm$ 0.9    &   $\pm$ 690     &    $\pm$ 0.05   &  $\pm$ 0.7     &  $\pm$ 0.42    \\
    \hline
    P7     &    37.3        &      842        &       1.05      &    56.6        &    92.7        \\
           &   $\pm$ 0.2    &   $\pm$ 99      &    $\pm$ 0.05   &  $\pm$ 0.7     &  $\pm$ 11.0    \\
    \hline
    G100   &    63.1        &      631        &       1.27      &    97.2        &    119         \\
           &   $\pm$ 0.5    &   $\pm$ 104     &    $\pm$ 0.06   &  $\pm$ 1.1     &  $\pm$ 20      \\
    \hline
    P25    &    37.1        &      491        &       1.05      &    64.3        &    165         \\
           &   $\pm$ 0.4    &   $\pm$ 34      &    $\pm$ 0.01   &  $\pm$ 0.2     &  $\pm$ 11      \\
    \hline
    G60    &    62.9        &      6.8        &       1.16      &    98.1        &    10700       \\
           &   $\pm$ 0.1    &   $\pm$ 0.2     &    $\pm$ 0.06   &  $\pm$ 0.9     &  $\pm$ 400     \\
           \hline
    \end{tabular}
    \label{tab:liquid_properties}
\end{table}

\subsubsection*{Equilibrium contact angle measurements}
Wetting equilibrium is difficult to identify on these systems using classical techniques such as droplet deposition. We see the contact line moving even a few hours after deposition. We circumvent this issue by deducing equilibrium contact angles $\theta_{eq}$ from experimental dynamic contact angles (Fig.\,\ref{fig:NLM}a): we fit the points around $U=0$ with a linear law, and take $\theta_{eq}$ equal to the value of the fitted contact angle at $U = 0$.

\subsubsection*{Gel preparation}
Gel slabs are prepared with a two-part commercial silicone kit (Dow Corning Sylgard 527). We mix equal volumes of each part of the kit together, as recommended by the manufacturer, in a weighing boat previously cleaned with ethanol and water, and dried. The gel mixture is degassed under vacuum for 2 \si{\hour} to remove bubbles. It is then poured in a $60 \times 40$-mm$^2$ plastic vessel (Caub\`{e}re), also cleaned with ethanol and distilled water and let to dry in a vacuum before use. Then, we leave the sample in an oven at \SI{65}{\celsius} for 15 to 18 \si{\hour}. We perform experiments exclusively on dust-free unmarked gels. 

\subsubsection*{Free-chain extraction}
We extract free chains from silicone gels using the process described in ref.\,\cite{hourlier-fargette2017}. After weighing pieces of gels, we dip them into toluene (VWR, AnalaR NORMAPUR), a good solvent for PDMS. Free chains migrate to the solvent. We renew toluene everyday for five days to accelerate the extraction process.  Then the sample is immersed in a mixture of toluene and ethanol (VWR, AnalaR NORMAPUR) to the bath to remove toluene from the gel. Ethanol is added progressively to avoid damaging the sample. We start with a solution of 20wt\% ethanol in toluene, and we increase the ethanol proportion by steps of 20wt\% every day, until the sample sits in 100\% ethanol. After three baths in pure ethanol, the gel stops shrinking. We dry the gel under vacuum to remove remaining solvent and weigh it again. From this procedure, we find that our materials contain $62$ wt\% free chains. This large amount of free chains explains likely the absence of transition between two sliding regimes \cite{hourlier-fargette2017,hourlier-fargette2018} in our experiments.

\subsubsection*{Rheology}
\begin{figure}[!ht]
    \centering
    \includegraphics[scale=1]{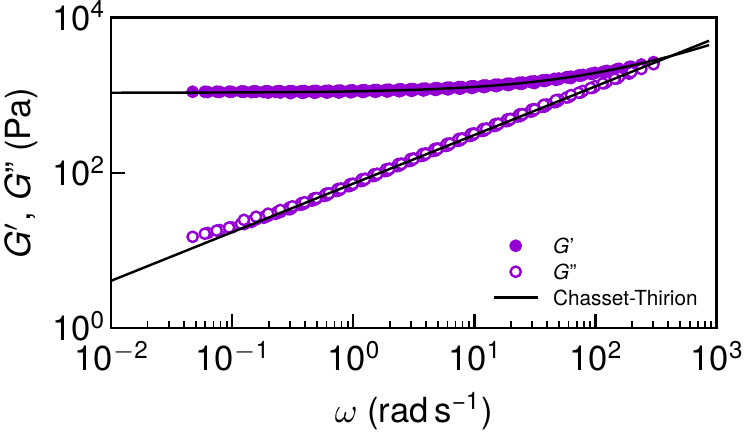}
    \caption{\textbf{Rheology of Sylgard 527.} Storage modulus $G'$ and loss modulus $G"$  as function of pulsation $\omega$. Strain amplitude: 1\%. Continuous black lines: Chasset-Thirion law, Eqs.\,\ref{eq:G'} and \ref{eq:G"}.}
    \label{fig:mat_met}
\end{figure}
We perform small amplitude oscillatory shear rheology on our gels with an Anton Paar MCR 501 rheometer mounted with a plate-plate geometry (diameter $d = 25$ mm). Strain amplitude is set at 1\%. We extend the frequency range using a time-temperature superposition procedure \cite{Rubinstein2003}. We fit the data with the Chasset-Thirion model \cite{curro1983}: 
\begin{equation}
    G(\omega)=\mu_0(1+(i\omega\tau)^m),
    \label{eq:CTlaw}
\end{equation}
with $G$ the complex modulus, $\omega$ the strain angular frequency and $\tau$ a characteristic relaxation time. The exponent $m$ depends on the degree of polymerization of the polymer network and on the coordination number of monomers \cite{curro1983,curro1985}.

If we decompose the complex modulus $G(\omega)$ in a real elastic part $G'(\omega)$ and an imaginary dissipative part $G"(\omega)$, we obtain
\begin{align}
G'(\omega) &= \mu_0(1 + \cos \left( m \pi/2 \right) (\omega \tau)^m),
\label{eq:G'}\\
G"(\omega) &= \mu_0 \sin \left(m \pi/2 \right) (\omega \tau)^m.
\label{eq:G"}
\end{align}
We determine $\tau$ and $m$ by fitting the loss factor deduced from the model to its experimental value:
\begin{equation}
\frac{G"}{G'} (\omega) = \tan{(\delta)} = \frac{\sin \left( m \pi/2 \right) (\omega \tau)^m \tan \left( m \pi/2 \right)}{\sin{\left( m \pi/2 \right)} (\omega \tau)^m + \tan{\left(m \pi/2 \right)}}.
\label{eq:loss_factor}
\end{equation}
We inject the values of $\tau$ and $m$ to fit equations \ref{eq:G'} and \ref{eq:G"} to the rheological data (Fig.\,\ref{fig:mat_met}). We find $\mu_0 = 1.077$ kPa, $\tau = 18.2$ ms, and $m = 0.626$.

\subsubsection*{Sliding experiments}
We deposit a liquid droplet with a micropipette on the gel. The spherical radius $R_0$ of the droplets before deposition is of the order of the capillary length $\ell_c=(\gamma/(\rho g))^{1/2}\sim\mathcal{O}(1.5)$ mm of the liquids in all cases.The experiment starts when we tilt the gel at an angle $\alpha$ with the horizontal. A LED panel (Effilux) shines light on the sample from below, and a camera (Imaging Source, DMK 33UX174) records top views of the droplet with a spatial resolution of 32 \si{\micro\meter\per\pixel}. We take side views on some experiments (Imaging Source, DMK 33UX174, spatial resolution 4 \si{\micro\meter\per\pixel}).  The thickness of all the samples, $h_{ s} \sim$ 4 mm, is much larger that the elastocapillary length of our material, $\ell_{ s}\sim 10$ \si{\micro\meter}, to avoid small-thickness effects \cite{zhao2018}. Samples are covered with a polystyrene lid that we find able to prevent surface ageing and dust deposition. We unmold gel layers and cut their edges so the meniscus is not in the way of side views. We obtain identical results when the silicone gel is in the box or unmolded.

We check the volume of droplets by weighing samples before droplet deposition and after. We tracked the motion of droplets with the software package FiJi \cite{schindelin2012a}. Droplet velocities range from $10^{-3}$ to 1 mm s$^{-1}$. In most cases, the trajectories that we observe are linear functions of time: droplets move at constant speed. For the longest experiments, drops may lose or gain water from surrounding air. In that case, we fit only the part of the trajectory that is unaffected. Thus we extract a single value $U$ of the droplet velocity from each experiment. Each set of experimental conditions is tested three times to ensure reproducibility.

\subsection{Derivation of scaling laws describing the dynamics of sliding droplets.}

\subsubsection{Vanishing values of the relaxation ratio $\mathcal{R}$}

\begin{figure*}[ht!]
    \centering
    \includegraphics[scale=1]{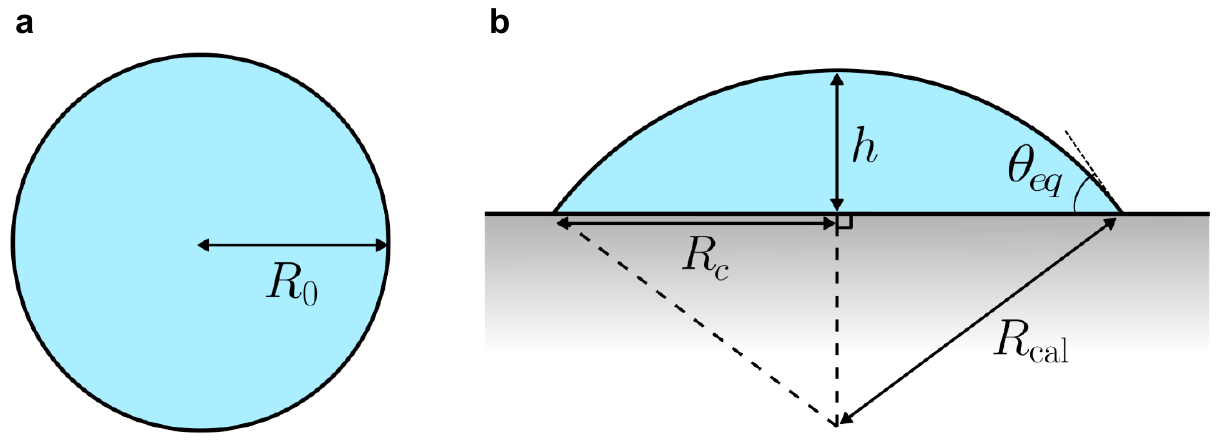}
    \caption{\textbf{Different configurations for a liquid drop. } (a) A spherical droplet of volume $V = 4 R_0^3/3 \pi$. (b) The same droplet as a spherical cap after spreading on a substrate, with radius of curvature  $R_{\rm cal}$, height at the center $h$, contact radius $R_c$ and contact angle $\theta_{eq}$.}
    \label{fig:calotte}
    \end{figure*}
 In experiments, a droplet of volume $V$ is deposited on the substrate. This volume  corresponds to a sphere of radius $R_0$ (Fig.\,\ref{fig:calotte}a). Once it has spread on the surface of the substrate, the droplet reaches its equilibrium shape, a spherical cap with radius $R_{\rm cal}$, height $h$, contact radius $R_c$ and contact angle $\theta_{eq}$ (Fig.\,\ref{fig:calotte}b).

 $R_c$ is difficult to measure in experiments, especially when the equilibrium contact angle is larger than $\pi/2$. In contrast, we can obtain $R_0$ by weighing the drop. If we know the equilibrium contact angle $\theta_{eq}$, we can write, assuming that droplets form spherical caps:
 
 \begin{equation}
 \left\{
 \begin{aligned}
     R_c  &=  R_{\mathrm{cal}} \, \cos\left( \frac{\pi}{2} - \theta_{eq} \right)\\
     R_{\mathrm{cal}} - h  &=  R_{\mathrm{cal}} \, \sin \left( \frac{\pi}{2} - \theta_{eq} \right).
 \end{aligned}
 \right.
 \end{equation} 
Rearranging,  we can link the contact radius and the height of the droplet to the spherical cap radius:
 \begin{equation}
 \left\{
 \begin{aligned}
     R_{c} &= R_{\rm cal} \sin \theta_{eq}\\
     h &= R_{\rm cal} (1 - \cos \theta_{eq}).
 \end{aligned}
 \right.
 \label{eq:hRc}
 \end{equation}
 Now, we can express the volume for both a sphere and a spherical cap:
 \begin{equation}
 \left\{
 \begin{aligned}
     V &= \frac{4 \pi}{3} R_0^3\\
     V &= \frac{\pi}{3} h^2 (3 R_{\rm cal} - h)
 \end{aligned}
 \right.
 \end{equation} 
 in terms of $R_c$ and $\theta_{eq}$:
 \begin{equation}
 \left\{
 \begin{aligned}
     V &= \frac{4 \pi}{3} R_0^3\\
     V &= \frac{\pi R_{c}^3}{3} \frac{(2 + \cos \theta_{eq}) (1 - \cos \theta_{eq})^2}{\sin^3 \theta_{eq}}
 \end{aligned}
 \right.
 \end{equation}
 Volume conservation then leads to:
  \begin{equation}
    \frac{R_c}{R_0}=\frac{1}{f(\theta_{eq})} = \sin \theta_{eq} \left( \frac{(2 + \cos \theta_{eq}) (1 - \cos \theta_{eq})^2}{4} \right)^{-\frac{1}{3}}.
    \label{eq:f_thetaEq}
 \end{equation}  

For a droplet sliding down an inclined plane, the ratio between its weight and the capillary force along its perimeter, called the Bond number, writes:
 
 \begin{equation}
 Bo_{\alpha} = \frac{\rho g R_0^3}{\gamma R_c} \sin \alpha. 
 \label{eq:defBondMat}
 \end{equation}

 Injecting Eq.\,\ref{eq:f_thetaEq} into Eq.\,\ref{eq:defBondMat}, we obtain an expression for the Bond number that accounts for changes in the equilibrium contact angle:
 
 \begin{equation}
 \boxed{
 Bo_{\alpha} = f(\theta_{eq}) \frac{\rho g R_0^2}{\gamma} \sin \alpha.
 }
 \label{eq:Bo_correc_theta}
 \end{equation}

Now, let's assume that the gravitational force experienced by the droplet is balanced dissipation in the liquid and contact angle hysteresis:
\begin{equation}
\rho g R_0^3 \sin \alpha \sim \eta \frac{U}{h} R_c^2+\gamma(\cos{\theta_a}-\cos{\theta_r})R_c.
\label{eq:gravLiqDiss}
\end{equation}
Here $U/h$ estimates the velocity gradient in the droplet, and $\theta_a$ and $\theta_r$ are the advancing and receding dynamic contact angles, \textit{i.e.} the threshold values of the contact angle above and below which contact line motion occurs. 

\begin{figure}[!htb]
    \centering
    \includegraphics[scale=.7]{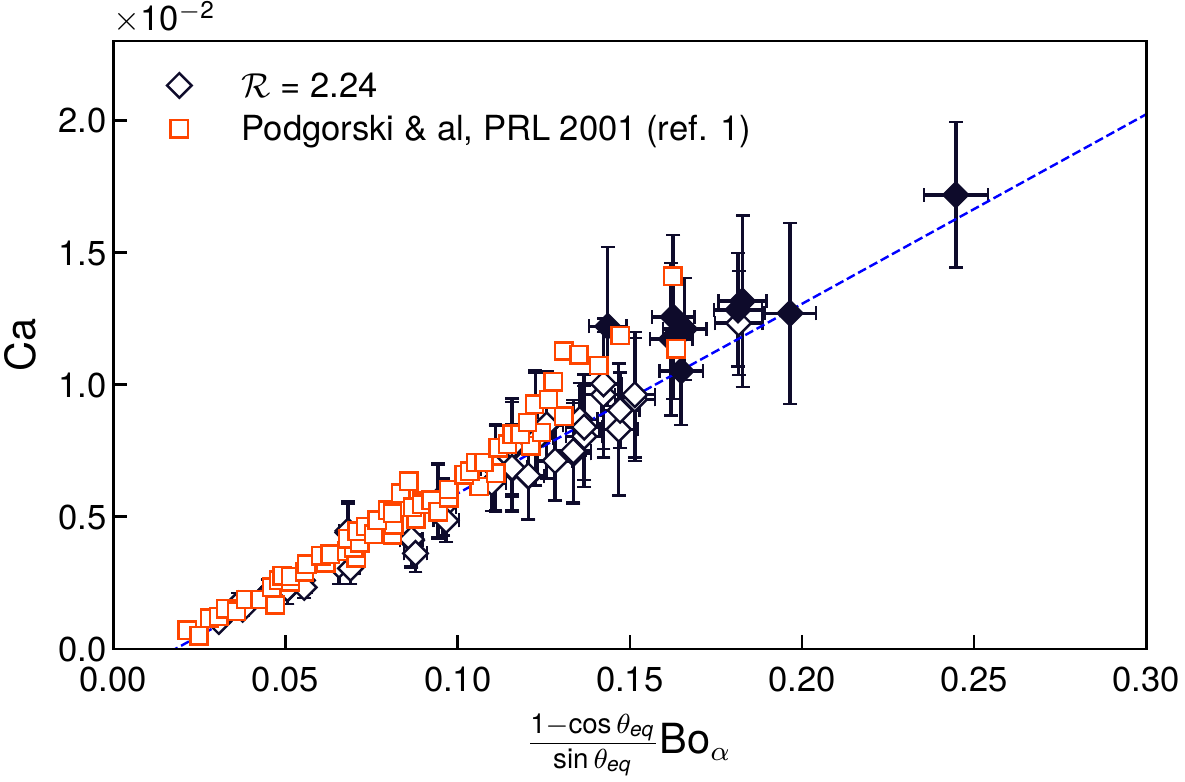}
    \caption{Comparison between our data for $\mathcal{R}\simeq 2$ and the data of Podgorski \textit{et al.\,} \cite{podgorski2001} once corrections related to a different equilibrium contact angle are accounted for.}
    \label{fig:podR2}
\end{figure}
In what follows, we assume that the values of dynamic contact angles are close to that of the equilibrium contact angle, a hypothesis that is valid at low droplet velocities, $Ca<<1$. Thus, we can use the description of the equilibrium shape of the droplet to relate the contact angle and the contact radius. Using Eq.\,\ref{eq:hRc}, we obtain:
\begin{equation}
    \frac{R_c^2}{h}=R_c\frac{\sin{\theta_{eq}}}{1-\cos{\theta_{eq}}}
\end{equation}
Dividing both sides by the liquid-vapor surface tension $\gamma$, replacing $R_c^2/h$, and using Eq.\,\ref{eq:f_thetaEq}, it follows that:
\begin{equation}
    Ca\sim \frac{1 - \cos{\theta_{eq}}}{\sin{\theta_{eq}}} [Bo_{\alpha}-Bo_c],
\label{eq:betterPod}
\end{equation}
where $Bo_c$ is a threshold Bond number below which contact hysteresis pins the droplet to the surface. This scaling was first suggested in ref.\,\cite{podgorski2001}. We observe that this correction leads to the overlap of our data for $\mathcal{R}\simeq 2$ and those of ref.\,\cite{podgorski2001} (Fig.\,\ref{fig:podR2}). Following Dussan \cite{dussanv.1985a} and Le Grand \textit{et al.} \cite{legrand2005}, we can evaluate contact angle hysteresis from the experimental value of $Bo_c$ using:
\begin{equation}
    Bo_c=\left(\frac{24}{\pi}\right)^{1/3}\frac{(\cos{\theta_r}-\cos{\theta_a})(1+\cos{\theta_a})^{1/2}}{(2+\cos{\theta_a})^{1/3}(1-\cos{\theta_a})^{1/6}}
\end{equation}
Le Grand \textit{et al.} performed this estimation accounting for the prefactor appearing in front of $Bo_{\alpha}$ when fitting their data with Eq.\,\ref{eq:betterPod}. We use the same procedure to obtain our estimate.

\subsubsection{Large-$\mathcal{R}$ sliding}
\label{subsec:largeRDyn}

We expect that the relation between injected energy and dissipation be modified when the effective viscosity of the solid exceeds that of the liquid. The collapse of the $Ca-Bo_{\alpha}$ curves in the $Ca_s-Bo_{\alpha}$ space, with $Ca_s=\mathcal{R}Ca$, suggests that the latter is a good metric. Besides, we know from the previous section that the equilibrium contact angle matters. Hence we should derive a scaling law for the solid-dominated case that accounts for all of these modifications. Note that we use again the approximation that the dynamic contact angles remain close to the equilibrium contact angle.  

Inspired by Ref.\,\cite{zhao2018}, we estimate the power dissipated per unit of volume of the solid  when the drop moves by a length $\ell$:
\begin{equation}
    d\mathcal{P}_{\rm diss}\sim \sigma \epsilon^2\omega
\end{equation}
with  $\sigma$ the viscous stress, $\epsilon$ the strain and $\omega$ the pulsation. Using the Chasset-Thirion model:
\begin{equation}
    G(\omega)=\mu_0(1+(i\omega\tau)^m),
    \label{eq:CTlaw}
\end{equation}
and taking the typical strain scale to be:
\begin{equation}
    \epsilon \sim \frac{\gamma}{\gamma_s} \sin{\theta_{eq}}
\end{equation}
and the characteristic pulsation of the experiment as:
\begin{equation}
    \omega =\frac{U}{\ell_s},
\end{equation}
we have the following estimate for viscous stresses in the solid:
\begin{equation}
    \sigma \sim \mu_0 \left(\frac{U \tau}{\ell_s}\right)^m.
\end{equation}
Then we have:
\begin{equation}
    d\mathcal{P}_{\rm diss}\sim\mu_0 \left(\frac{U \tau}{\ell_s}\right)^m\left(\frac{\gamma}{\gamma_s} \sin{\theta_{eq}}\right)^2\frac{U}{\ell_s}\\
\end{equation}
Dissipation takes place in a half-torus having a radius $R_c$, width $\ell$ and height $\ell_r$. As elasticity balances the vertical component of the resulting capillary force per unit length at the contact line $\gamma \sin{(\theta_{eq})}$, the height of the ridge scales as:
\begin{equation}
    \ell_r \sim \gamma \sin{\theta_{eq}}/\mu_0.
\end{equation}
Then, we can estimate the power dissipated in the solid, neglecting numerical prefactors:
\begin{align}
&\mathcal{P}_{\rm diss} \sim \sigma \epsilon^2 \omega R_c \ell_r \ell\\
&\mathcal{P}_{\rm diss} \sim \mu_0 U R_c\ell \left(\frac{\gamma}{\gamma_s} \sin{\theta_{eq}}\right)^3\left(\frac{U \tau}{\ell_s}\right)^m
\end{align}

Now, we can write the force balance that a droplet sliding on a viscoelastic substrate should obey. Dividing $\mathcal{P}_{\rm diss}$ by the sliding velocity $U$, we have:
\begin{equation}
    \rho g R_0^3\sin{\alpha}\sim\mu_0 R_c\ell \left(\frac{\gamma}{\gamma_s} \sin{\theta_{eq}}\right)^3\left(\frac{U \tau}{\ell_s}\right)^m.
\end{equation}
Dividing by the liquid-vapor surface tension $\gamma$ on both sides and rearranging, we obtain:
\begin{equation}
    Bo_{\alpha} \sim \frac{\mu_0\ell}{\gamma}\left(\frac{\gamma}{\gamma_s} \sin{\theta_{eq}}\right)^3\left(\frac{U \tau}{\ell_s}\right)^m
\end{equation}
and using:
\begin{equation}
    \frac{U\tau}{\ell_s}=\mathcal{R}Ca=Ca_s,
\end{equation}
we end up with the following prediction:
\begin{equation}
    {Bo_{\alpha}\sim \frac{\mu_0\ell}{\gamma}\left(\frac{\gamma}{\gamma_s} \sin{\theta_{eq}}\right)^3{Ca_s}^m}.
    \label{eq:scalingSolidDiss}
\end{equation}
The prefactor $[(\gamma/\gamma_s) \sin {(\theta_{eq})}]^3$ in Eq. \ref{eq:scalingSolidDiss} should capture the dependence on equilibrium contact angles. In the limit of thick samples,  $\ell=\ell_s=\frac{\gamma_s}{2\mu_0}$ \cite{zhao2018}, and  we obtain Eq.\,7 in the main text when neglecting numerical prefactors:
\begin{equation}
    {Bo_{\alpha}} \sim \left( \frac{\gamma}{\gamma_{ s}} \right)^2 (\sin{\theta_{eq}})^3 {Ca_s}^m.
    \label{eq:solid_scaling}
    \end{equation}

\subsection*{Comparison of experimental $Ca-Bo_{\alpha}$ curves with the non-linear model}

\begin{figure*}[!htb]
    \centering
    \includegraphics[scale=1]{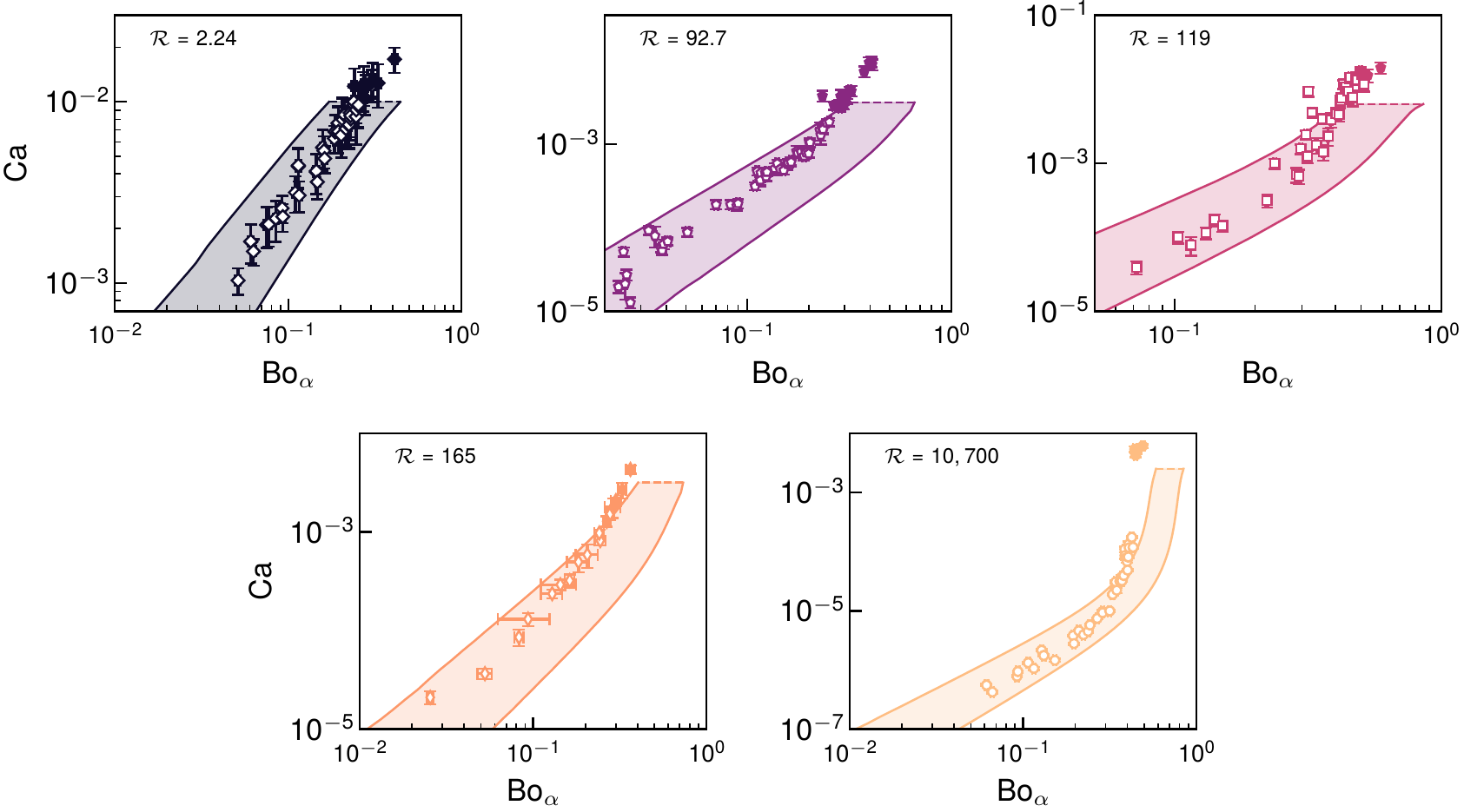}
    \caption{Comparison between the data of Fig.\,2 in the main text with the non-linear model of Dervaux \textit{et al.\,}\cite{dervaux2020}. Closed symbols represent data points obtained beyond the pearling threshold. The fitting domains represent the variation of the fitting results due to uncertainty on the experimental parameters. Dashed lines: pearling threshold capillary number.}
    \label{fig:figS2}
\end{figure*}

Figure \ref{fig:figS2} shows a comparison between experimental data for the $Ca-Bo_{\alpha}$ displayed in Fig.\,2 in the main text and theoretical predictions from our non-linear model \cite{dervaux2020}. Theoretical fits cover a surface in the $Ca-Bo_{\alpha}$ space because we account for uncertainties in experimental parameters such as the viscosity and the surface tension of the liquids, etc. Fits are seen to be in good agreement with experimental data. 

\end{document}